\documentclass[12pt]{iopart}

\usepackage{graphicx}
\usepackage{bm}
\usepackage{iopams}
\usepackage{xcolor}

\def\fetese{FeSe$_{1-x}$Te$_x$}

\begin{document}


\title{  Nematic properties of FeSe$_{1-x}$Te$_{x}$ crystals with a low Te content  }

\author{Y.A.~Ovchenkov$^{1*}$, D.A.~Chareev$^{2,3.4}$, V.A.~Kulbachinskii$^{1,5}$, V. G.~Kytin$^{1,6}$,  D. E.~Presnov$^{7,8}$, Y.~Skourski$^9$, L.V.~Shvanskaya$^{1,10}$, O.S.~Volkova$^{1,3,10}$,  D.V.~Efremov$^{11**}$  and A.N.~Vasiliev$^{1,3,12}$}

\address{$^1$Lomonosov Moscow State University, Moscow 119991, Russia}

\address{$^2$Institute of Experimental Mineralogy, RAS, Chernogolovka, 142432, Russia}
\address{$^3$Ural Federal University, Ekaterinburg  620002 , Russia}
\address{$^4$Kazan Federal University, Kazan 420008 Russia}
\address{$^5$Moscow Institute of Physics and Technology, Dolgoprudny 141700, Russia}
\address{$^6$VNIIFTRI, Mendeleevo, Moscow region 141570, Russia}
\address{$^7$Skobeltsyn Institute of Nuclear Physics, Moscow 119991, Russia}
\address{$^8$Quantum Technology Centre, Lomonosov Moscow State University, Moscow 119991, Russia}
\address{$^{9}$Dresden High Magnetic Field Laboratory (HLD-EMFL), HZDR,  Dresden, Germany}
\address{$^10$National University of Science and Technology 'MISiS' , Moscow 119049, Russia}
\address{$^{11}$IFW-Dresden, Institute for Solid State Research, D-01171 Dresden, Germany}
\address{$^{12}$National Research South Ural State University, Chelyabinsk 454080 , Russia}

\ead{$^*$ovtchenkov@mig.phys.msu.ru, $^{**}$d.efremov@ifw-dresden.de}

\date{\today}
%
\begin{abstract}

We report on the synthesis and physical properties of FeSe$_{1-x}$Te$_x$ single crystals with a low Te content (x = 0.17, 0.21, 0.25), where the replacement of Se with Te partially suppresses superconductivity. Resistivity and Hall effect measurements indicate weak anomalies at elevated temperatures ascribed to nematic transitions. A quasi-classical analysis of transport data, including in a pulsed magnetic field of up to 25 T, confirms the inversion of majority carriers type from holes in FeSe to electrons in FeSe$_{1-x}$Te$_x$ at x $>$ 0.17. The temperature-dependent term in the elastoresistance of the studied compositions has a negative sign, which means that for substituted FeSe compositions, the elastoresistance is positive for hole-doped materials and negative for electron-doped materials just like in semiconductors such as silicon and germanium.   

%
\end{abstract}  

\maketitle

\section{Introduction}

FeSe is one of the most intriguing compounds among iron-based superconductors (IBS), first of all because
it undergoes a transition to the nematic phase at $T_s = 90$K (structural transition from the tetragonal at ambient conditions to the orthorhombic low temperature phase)  as well as to superconductivity at $T_c=9  $K, and shows no evidence of a striped antiferromagnetic phase (AFM) \cite{Johnston2010,Bohmer2018}. Furthermore, the NMR data  at $T_s$ are consistent with an uncorrelated Fermi liquid. The AFM spin fluctuations (SFs) are strongly enhanced only below $T_s$ \cite{Baek2015,Boehmer2015,Baek2016}.  
It makes it very different from  other IBS, in which the structural transition is followed by the transition to the striped AFM phase.
Under pressure the superconducting critical temperature increases from $T_c=9$~K at ambient pressure to $\sim 37$~K at $P=6$~GPa \cite{Terashima2015,Miyoshi2014,Kothapalli2016}. Simultaneously the structural transition temperature decreases from $T_s=90$~K to $\sim20$~K in this pressure range \cite{Terashima2015,Miyoshi2014,Kothapalli2016}. In addition, the AFM state emerges for $P>0.8$GPa \cite{Bendele2010,Bendele2012,Terashima2016,Kaluarachchi2016,Hosoi2016}. These observations suggest strong competition between the magnetic and nematic phases and their importance for superconductivity.

Substitution of sulfur for selenium suppresses the low temperature nematic order in FeSe  \cite{Ovch_JLTP, SUST-30-3-035017, Char_FeSeS_CEC} that is accompanied with many anomalous changes in properties including the strong changes in a superconducting pairing which was clearly observed by a tunnel microscopy \cite{hanaguri2017}.
Substitution of Te on the Se site  is expected to exert negative chemical pressure. But so far only  FeSe$_{1-x}$Te$_x$ single crystals with doping $x>0.3$ were investigated. At $x=0.3$ dopings no nematic phase was observed for all range of temperatures. To complete the phase diagram of the nematic phase of FeSe one need high quality crystals with doping $x<0.3$.

Here we present a method of the synthesis which allow to grow high quality single crystals of FeSe$_{1-x}$Te$_x$ with a low Te content, and discuss the properties of  FeSe${}_{1-x}$Te${}_{x}$  crystals with x=0.17, 0.21, and 0.25 which shed light on the origin of the nematic state in the FeSe family.
The paper is constructed as following. We start with the description of the sample preparation and its characterization. Then we discuss the magnetotransport data obtained on these crystals and present the data on their elastoresistivity. Finally, all the results are summarized in the Discussion and Conclusion  sections.

\section{Sample preparation and experiments}

\begin{table*}[ht]
  \caption{\label{tbl:T1} Summary of the studied FeSe${}_{1-x}$Te${}_{x}$ samples properties. The value of x$_{L}$ is the Te content in the FeSe${}_{1-x}$Te${}_{x}$ precursors load. The value of x$_{EDS}$ is the Te content in the crystals determined by an energy-dispersive spectroscopy.  $T_{c}$ is the superconducting critical temperature determined from $\rho_{xx}(T)$. The mematic transition critical temperature $T_{N}$ is determined from the temperature dependence of $d\rho_{xx}/dT$. The parameters $a$, $c$, and $V$ are the lattice constants and cell volume respectively (for FeSe, this parameters are taken from Ref.\cite{CrystEngComm12.1989})}
  \begin{tabular}{cccccccc}
    \br
    Sample & $x{}_{L}$ & $x{}_{EDS}$ & $T_{c}$ (K) &  $T_{N}$ (K) & $a$ (\r{A}) & $c$ (\r{A}) & $V$  (\r{A}$^{3}$)\\
    \mr
    FeSe  &  & & 9.1 & 85 &3.765 & 5.518& 78.2(2)  \\
    FeSe${}_{1-x}$Te${}_{x}$  & 0.10 & 0.17$\pm$0.01 &7.5 - 8.2& 70 &3.788(4) &5.671(7) & 81.37(15)  \\
    FeSe${}_{1-x}$Te${}_{x}$ & 0.20 & 0.21$\pm$0.01 &5.2 - 6.3 & 60 &3.784(2) &5.727(6) & 82.9(6)  \\
    FeSe${}_{1-x}$Te${}_{x}$ & 0.25 & 0.25$\pm$0.01 & 6.5& 60 &3.795(6) &5.778(14) & 83.2(3)  \\

    \br
  \end{tabular}
\end{table*}

The tetragonal FeSe${}_{1-\delta}$ phase coexists with a pure iron or a hexagonal Fe${}_{7}$Se${}_{8}$ phase \cite{okamoto1991fe}. The peritectoid reaction FeSe${}_{1-\delta}$ = Fe${}_{7}$Se${}_{8}$ + Fe of a tetragonal phase decomposition occurs at temperatures above 457~$^{\circ}$C. The homogeneity range of a tetragonal phase is from FeSe${}_{0.96}$ to FeSe${}_{0.975}$ \cite{gr1968heat}.
Phase relationships for mixing of FeSe and FeTe are not well known. Mixing FeSe and FeTe in any proportion leads to a formation of the solid solution but it is likely that there is a two-phase region or a spinodal near 50\%. Taking into account that FeTe decomposes at 800~$^{\circ}$C, and FeSe decomposes at 457~$^{\circ}$C, it is easy to understand the reduction of the temperature of stability of Fe(Te,Se) as the selenium content increases. Therefore, the synthesis temperature of tetragonal Fe(Se,Te)${}_{1-\delta}$ solid solutions should be chosen according to selenium and tellurium ratio in the sample. In flux method, to select the proper synthesis temperature, the solubility and diffusion of tellurium should be also taken into account. Our previous experiments show that at temperatures below 450~$^{\circ}$C in the AlCl${}_{3}$/KCl/NaCl eutectic mixture, the most common solution for FeSe and Fe(SeS) growth, the transfer of Te occurs very weakly and often non-uniformly, probably because of a very weak solubility and diffusion of tellurium. Therefore, for obtaining high-quality crystals with a low tellurium content, we used the synthesis temperatures from the range slightly above the decomposition point of a tetragonal FeSe.

The series of FeSe${}_{1-x}$Te${}_{x}$ single crystals was prepared using the AlCl${}_{3}$/KCl/NaCl eutectic mixture in evacuated quartz ampoules  in a constant temperature gradient \cite{CrystEngComm12.1989}. Quartz ampoules with the Fe(Te,Se) charge and the maximum possible amount of the AlCl${}_{3}$/KCl/NaCl eutectic mixture were placed in the furnace with their hot ends at a temperature of $\sim$535~$^{\circ}$C and their cold ends at a temperature of $\sim$453~$^{\circ}$C. The chalcogenide charge gradually dissolved in the hot end of the ampoules and precipitated as single crystals at the cold end. After eight weeks of heating, iron monochalcogenide plate-like crystals were found at the cold ends of the ampoules.

The chemical composition of the grown crystals was determined using a Tescan Vega II XMU scanning electron microscope equipped with an INCA Energy 450 energy-dispersive spectrometer; the accelerating voltage was 20 kV. The determination of the unit cell parameters was performed by an automatic procedure on a four-circle Oxford Diffraction Xcalibur diffractometer (graphite-monochromated Mo-K$\alpha$ radiation, $\lambda$=0.71073 \r{A}) equipped with an Xcalibur S-area detector. The exposure time varied from 4 to 8 sec depending on crystal size. For unit cell determination at least 4-5 crystals of a square plate-like morphology only were studied from each batch. The initial chemical composition of batches, chemical composition of crystals obtained and their crystallographic characteristics along with the properties of pure FeSe are listed in Table~\ref{tbl:T1}.

 Quasiclassical multicarrier (three-band) analysis of the field dependencies of the conductivity tensor components is used to extract the carriers mobilities and concentrations. The details of the used analysis method are published elsewhere \cite{SUST-30-3-035017}. Elastoresistivity was measured using a method similar to that described in Ref. \cite{Chu710} using a commercial piezoelectric device and AC transport option of a Quantum Design PPMS system equipped with a multifunctional insert MF-130.


\section{Results}

\begin{figure}[h]
\centering
  \includegraphics[scale=0.5]{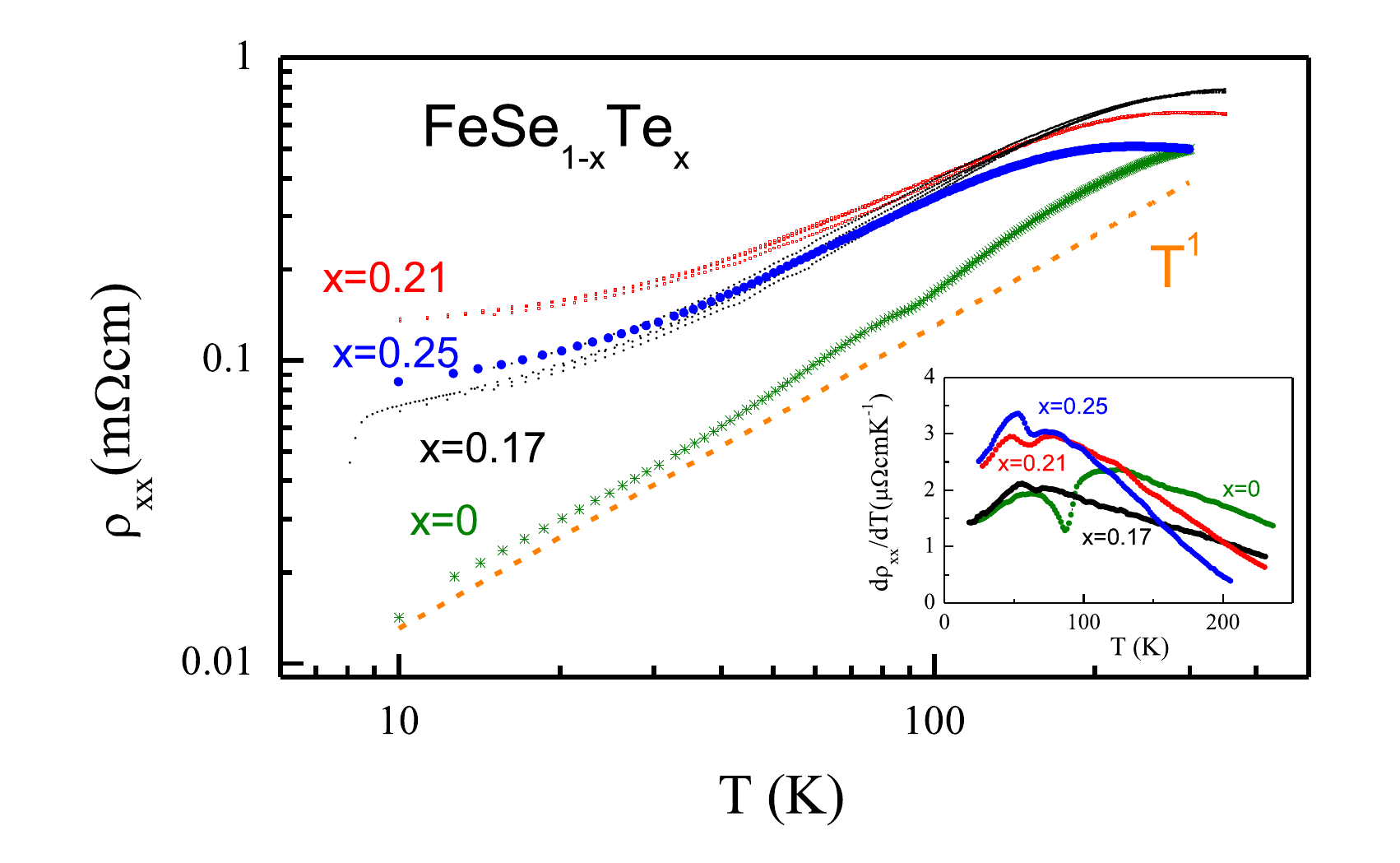}
  \caption{Log-log plot of the temperature dependence of $\rho_{xx}$ for \fetese at $x=0$ (from Ref. \cite{2017ovchenkovMISM}), $x=0.17$, $x=0.21$, and $x=0.25$. The straight dashed line shows a linear dependence.  Inset: $d\rho_{xx}/dT$ for selected crystals. }
  \label{fgr:fig1}
\end{figure}

Pure tetragonal crystals, suitable for further transport investigation, were found in three batches hereinafter referred to as the $x=0.17$, $x=0.21$,  and $x=0.25$ batches according to Te content in crystals. A detailed discussions of the quality of the obtained crystals, their superconducting properties and used method of extraction of the carriers band parameters are given in the supplementary materials. Here we present the results confirming the nematic transitions in the studied crystals and results of the quaisiclassical analysis of their electronic properties. Fig.~\ref{fgr:fig1} shows the temperature dependence of the resistivity $\rho_{xx}(T)$. The data for the crystals with dopings $x=0.17$ and $x=0.21$ are normalized to the mean value for each batch at $T=250$K. For comparison we add $\rho_{xx}(T)$ of undoped FeSe studied in Ref. \cite{2017ovchenkovMISM}.  For crystals from the same batch, the curves $\rho_{xx}$(T) almost coincide, that shows a good homogeneity of the crystals in batches.  All the curves are relatively smooth  and show similar behavior. The inset of Fig.\ref{fgr:fig1} presents temperature derivatives of resistivity curves. For doped samples $d\rho_{xx}/dT$ has a feature in the form of a local minimum similar to that at $T=90$K for pure FeSe but it occurs at lower temperatures.

\begin{figure}[h]
\centering
  \includegraphics[scale=0.5]{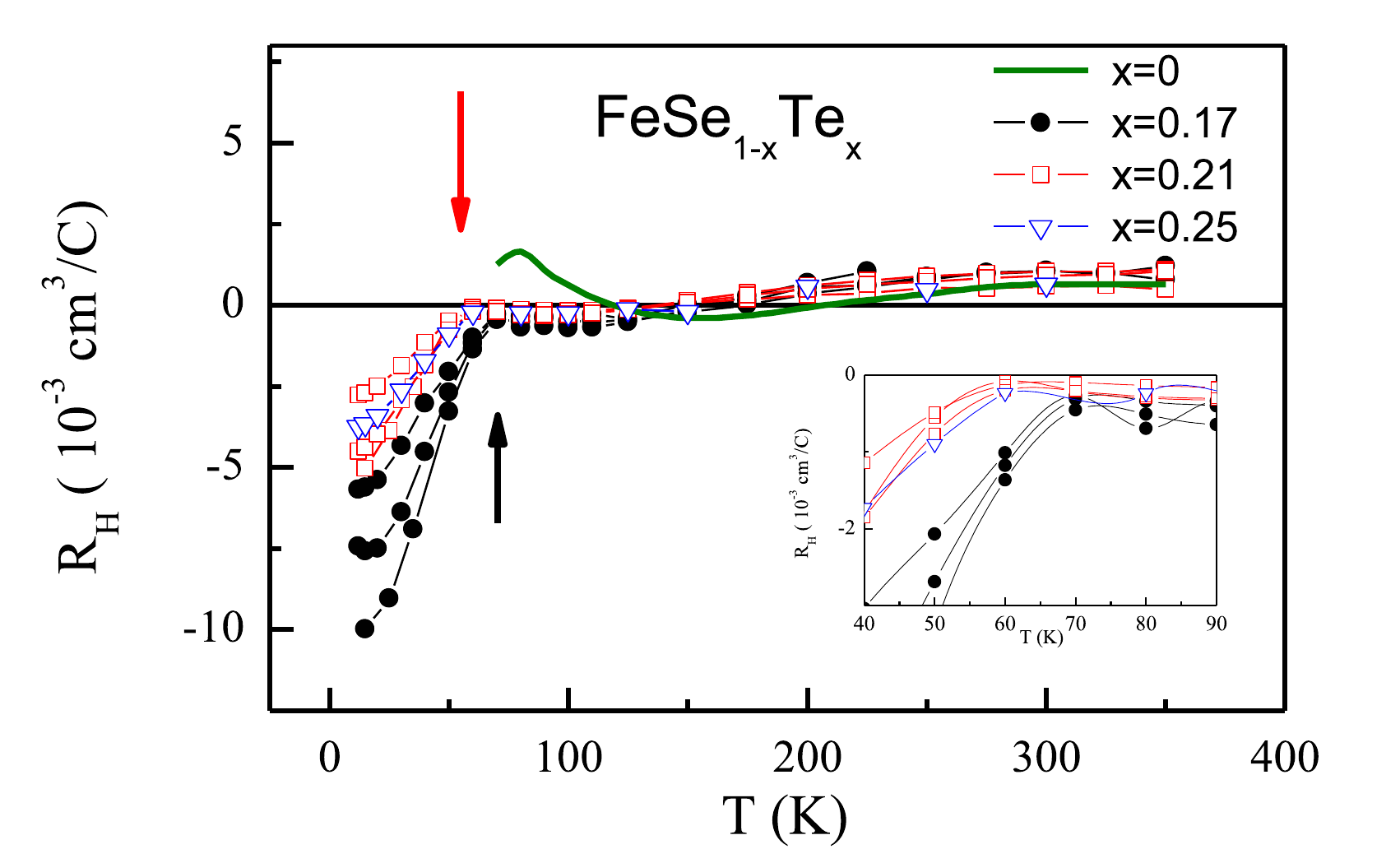}
  \caption{ Temperature dependence of the Hall coefficient for studied \fetese crystals and for the reference FeSe crystal.}
  \label{fgr:fig2}
\end{figure}

The Fig.\ref{fgr:fig2} shows the temperature dependence of the Hall coefficient $R_{H}(T)$ for doped \fetese  and for the reference undoped FeSe crystal studied in Ref. \cite{2017ovchenkovMISM}. At high temperatures, compounds show weak temperature dependence of Hall coefficient. At temperatures $T_{L}=70$~K for $x = 0.17$ and $T_{L} =60$~K  for $x=0.21$ and $x=0.25$ the Hall coefficients have kinks. With lowering of the temperature  the Hall coefficients strongly increases in absolute values, suggesting a Lifshitz transition at these temperatures. The sign of the Hall coefficients indicates that new electron pockets emerge.

\begin{figure}[ht]
\centering
  \includegraphics[scale=0.5]{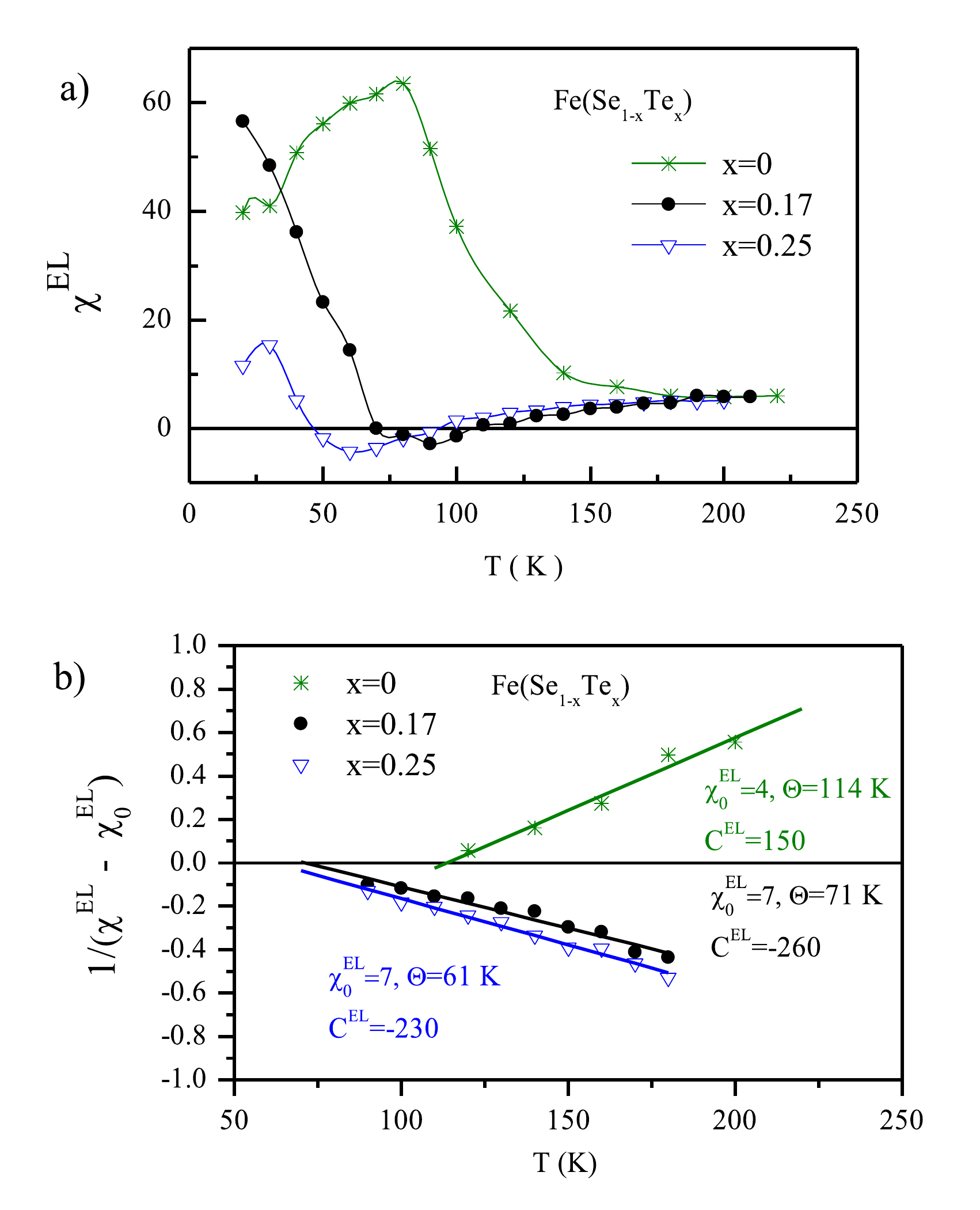}
  \caption{(a)Temperature dependence of the longitudinal elastoresistivity $\chi^{EL}=(\Delta{}\rho_{xx}/\rho_{xx})/(\Delta{}l_{x}/l_{x})$ (see in text) for \fetese x=0.17, x=0.25 and undoped FeSe crystals. (b) Temperature dependence of the inverse of $(\chi^{EL}-\chi^{EL}_{0})$. Data are fitted with the Curie-Weiss type dependence $\chi^{EL}~=~{C^{EL}}/{(T-\theta})~+~\chi^{EL}_{0}$. }
  \label{fgr:fig-3}
\end{figure}

The Fig.\ref{fgr:fig-3}(a) shows temperature dependence of  the longitudinal elastoresistivity $\chi^{EL}=(\Delta{}\rho_{xx}/\rho_{xx})/(\Delta{}l_{x}/l_{x})$ determined as a ratio of the change in longitudinal resistivity $(\Delta{}\rho_{xx}/\rho_{xx})$ to the change in the sample length $(\Delta{}l_{x}/l_{x})$ under applied deformation. Two \fetese crystals and one pure FeSe crystal were measured under the same conditions. First, despite the fundamental difference in the temperature behavior of  the longitudinal elastoresistivity, it saturates at high temperatures at the same value near about 6 which is still far from the gage factor of the ordinary metals. Thus, the temperature independent part of the elastoresistivity may have a common mechanism for all the studied compositions. Another common feature of all curves is a local extrema between 50 and 100 Kelvin. For FeSe the maximum of elastoresistivity occurs at the nematic transition. We consider that for other compounds the extrema  may also coincide with the temperatures of nematic transitions.

In the Fig.\ref{fgr:fig-3}(b) we plot the Curie-Weiss type fit for the temperature dependence of the measured elastoresistivity. The Curie-Weiss dependence describes well our experimental data using reasonable values of the parameters. The value of $\chi^{EL}_{0}$ is the experimentally measured residual elastoresistivity at high temperature. The values of $\theta$ are in a good agreement with the temperatures of the observed anomalous points in resistivity and Hall coefficients. The main difference between \fetese and FeSe is a negative sign of the coefficient $C^{EL}$ of the Curie-Weiss dependence for the former compounds.   

Thus, the temperature-dependent part of the elastoresistivity of the studied compositions can be described by the same type of dependence, which is usually used for FeSe. This leaves no doubt about the nematic nature of the \fetese under study.

To study the carriers properties in \fetese we applied the quasiclassical multi-carrier analysis to the experimental magnetotransport data. The experimental data and details of analysis we describe in supplementary materials section.  The similar method reviled a strong increase of the carriers concentration in FeSe with sulfur substitution \cite{ovch_invers} which is in good agreement with results of a quantum oscillation study of this series \cite{Coldea2019}.

The mobility of the main carriers in \fetese is substantionaly lower than in FeSe. This means that for the same accuracy in determining the parameters, significantly higher values of magnetic fields are required. To achieve the required accuracy, we carried out measurements of the magnetotransport properties of compositions with x = 0.17 in pulsed fields at low temperatures. The optimal relation  between the induced noise and the field range value for the studied samples was reached at about 25 Tesla. The results of the multi-carrier analysis of the transport data in DC and pulsed magnetic fields are listed in  Table~\ref{tbl:T2}. The summary of the electronic properties of our \fetese samples and FeSe${}_{1-x}$S${}_{x}$ samples  studied in \cite{ovch_invers} are plotted in Fig.\ref{fgr:fig-5}.

\section{Discussion}

The FeSe family exhibits many unusual features in the electronic properties \cite{2017_Coldea}, some of which are related to the nematic nature of these compounds. The nematicity is intensively studied in FeSe${}_{1-x}$S${}_{x}$ series. To our knowledge, our work is the first report on nematicity in \fetese. Some signs, such as a decrease in transition temperature, indicate that these compounds are on the other side of the ``nematic dome''. Therefore, it is particularly interesting to find a change in the sign of the coefficient of elastoresistivity $C^{EL}$ and the majority carrier type inversion which apparently occur simultaneously. 
This phenomenon may be the key to understanding the nematicity of FeSe. In addition, the detected sign-reversal of the elastoresistivity in FeSe family, as shown below, can have a common cause with the sign-reversal of the in-plane anisotropy of resistivity in hole-doped compounds of BaFe${}_{2}$As${}_{2}$ series \cite{blomberg2013sign}

Pure iron selenide and its compounds with isovalent substitutions are theoretically fully compensated semimetals. Real synthesized compounds usually have a sufficient degree of non-stocheometry to expect a violation of the compensation by units or even tens of percent. The study of the magnetotransport properties of FeSe${}_{1-x}$S${}_{x}$ series revealed a systematic change in the ratio of electrons and holes \cite{ovch_invers} with possible inversion of the type of majority carriers in FeSe${}_{1-x}$Te${}_{x}$ series, which is confirmed in the current study.

The origin of the carrier-type inversion in FeSe family is not clear and needs further study. The angle-resolved photoemission spectroscopy (ARPES) study of the strained FeSe films \cite{PhysRevB.95.224507} showed that under tensile strain the Fermi level moves close to the Van Hove singularity in the tetragonal state and the Lifshitz transition occurs in the nematic state. This study provides a possible microscopic scenario of the majority carriers type inversion in FeSe family. It is expected that the investigated substitutions of selenium cause a similar lattice deformation. We suppose that in the region of tellurium concentrations approximately less than 15\% there can exist a quantum critical point corresponding to the coincidence of the Fermi level with the Van Hove singularity. This assumption is consistent with the results of muon spin rotation experiments which find the proximity of FeSe to quantum criticality \cite{PhysRevB.97.201102} .

On the other hand the change in the electronic properties of bulk samples can  be a simple consequence of the variation of the synthesis conditions. For example, the temperature of synthesis for FeSe${}_{1-x}$Te${}_{x}$ is normally much higher than for FeSe${}_{1-x}$S${}_{x}$ which can yield other type of structural defects. Regardless of the microscopic reasons, our study indicates that the synthesized  FeSe${}_{1-x}$Te${}_{x}$ with low Te content are nematic compounds with opposite sign of majority carriers as compared to FeSe.

\begin{figure}[h]
\centering
  \includegraphics[scale=0.5]{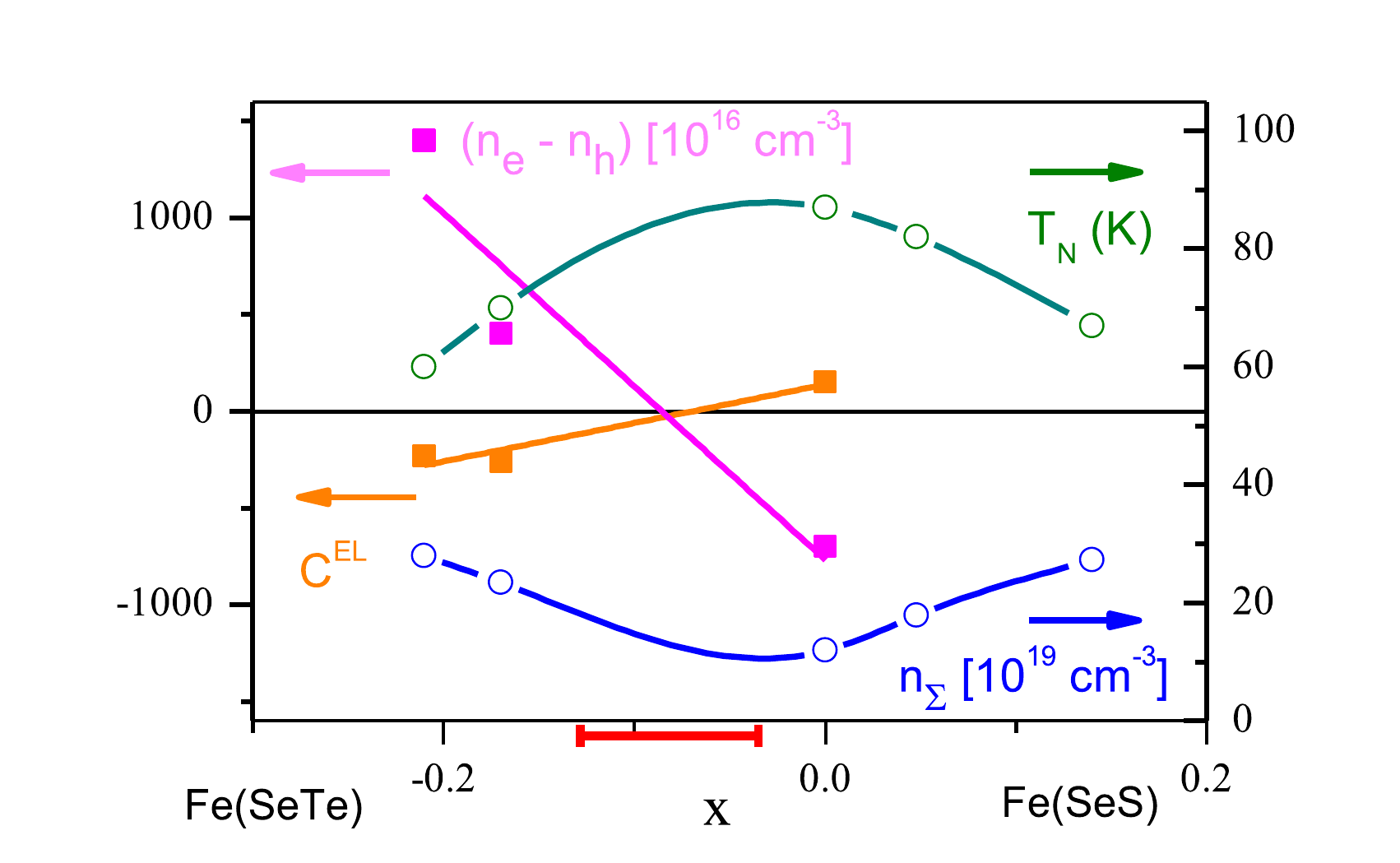}
  \caption{Dependence of the structural transition temperature $T_{N}$, overall carriers consantration $n_{\sum}=n_{electron}+n_{hall}$ in nematic state at low temperature and longitudinal elastoresistivity constant $C^{EL}$ on the chemical pressure parameter $x$. Positive x stand for compositions with sulfur and negative x are for \fetese. (Data for compositions with sulfur are taken from Ref. \cite{ovch_invers}) }
  \label{fgr:fig-5}
\end{figure}

\begin{figure}[h]
\centering
  \includegraphics[scale=0.5]{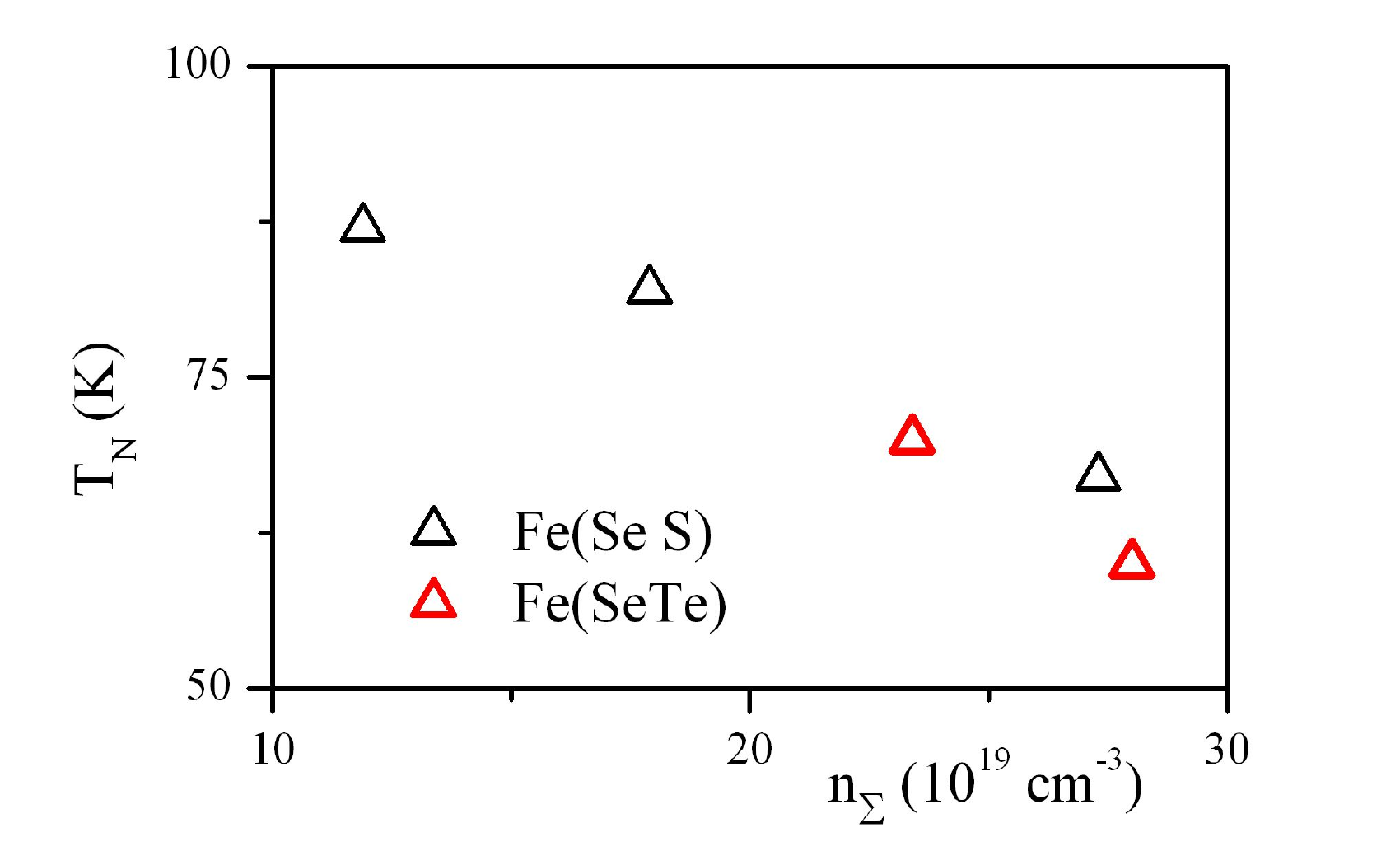}
  \caption{ Relation between the structural transition temperature $T_{N}$ and overall carriers consantration $n_{\sum}=n_{electron}+n_{hall}$ for FeSe${}_{1-x}$S${}_{x}$ and \fetese. (Data for compositions with sulfur are taken from Ref. \cite{ovch_invers}) }
  \label{fgr:fig-5_2}
\end{figure}

The Fig.\ref{fgr:fig-5} demonstrates that for the FeSe series the local minimum in the total carrier concentration coincides with the local maximum of the nematic transition temperature, which is consistent with the electronic nature of this phase transition. The Fig.\ref{fgr:fig-5_2} shows the relationship between structural transition temperature and carrier concentration in more detail.

%

\begin{table*}[ht]
\small
  \caption{\ Results of the simultaneous fitting of $\sigma_{xy}(B)$ and $\sigma_{xx}(B)$ with three-band model in field range up to 7 T  for  FeSe${}_{0.79}$Te${}_{0.21}$ and in field range up to 25 T for FeSe${}_{0.83}$Te${}_{0.17}$}
  \label{tbl:T2}
  \begin{tabular*}{\textwidth}{@{\extracolsep{\fill}}lllllll}
    \hline
    Sample &  $n_{h}$ & $\mu_{h}$  &  $n_{e1}$ & $\mu_{e1}$ &  $n_{e2}$ & $\mu_{e2}$ \\
			&(10$^{19}$ cm$^{-3}$) & (cm$^{2}$/Vs) & (10$^{19}$ cm$^{-3}$) & (cm$^{2}$/Vs)&(10$^{19}$ cm$^{-3}$) & (cm$^{2}$/Vs) \\
    \hline
    FeSe${}_{0.83}$Te${}_{0.17}$ & 11.5 &303 &11.9 & 343 & 0.32 & 1830\\
    FeSe${}_{0.79}$Te${}_{0.21}$ & 13.3 &152 & 14.7 & 163 & 0.11 & 890 \\
    \hline
  \end{tabular*}
\end{table*}

\section{Conclusion}

The study of the nematic properties of FeSe$_{1-x}$Te$_{x}$ compounds weth a low Te content provides new important experimental materials about elastoresistivity and other electronic properties of the iron-based superconductor,

\section{Acknowledgments}
This work has been supported by the Ministry of Education and Science of the Russian Federation in the framework of Increase Competitiveness Program of NUST “MISiS” (project K2-2017-084) and Kazan Federal University, by Act 211 of Government of the Russian Federation, contracts 02.A03.21.0004, 02.A03.21.0006 and 02.A03.21.0011. We acknowledge support of Russian Foundation for Basic Research through Grants 16-29-03266 and 17-29-10007. We also acknowledge the support of HLD at HZDR, a member of the European Magnetic Field Laboratory (EMFL).

\section*{References}
 \bibliographystyle{unsrt} 
 \bibliography{FeSeTe_low}

\section{SUPPLEMENTARY MATERIAL}

\renewcommand{\thefigure}{S\arabic{figure}}

\setcounter{figure}{0}

\subsection{Samples characterization}

\begin{figure}[h]
\centering
 \includegraphics[height=5cm]{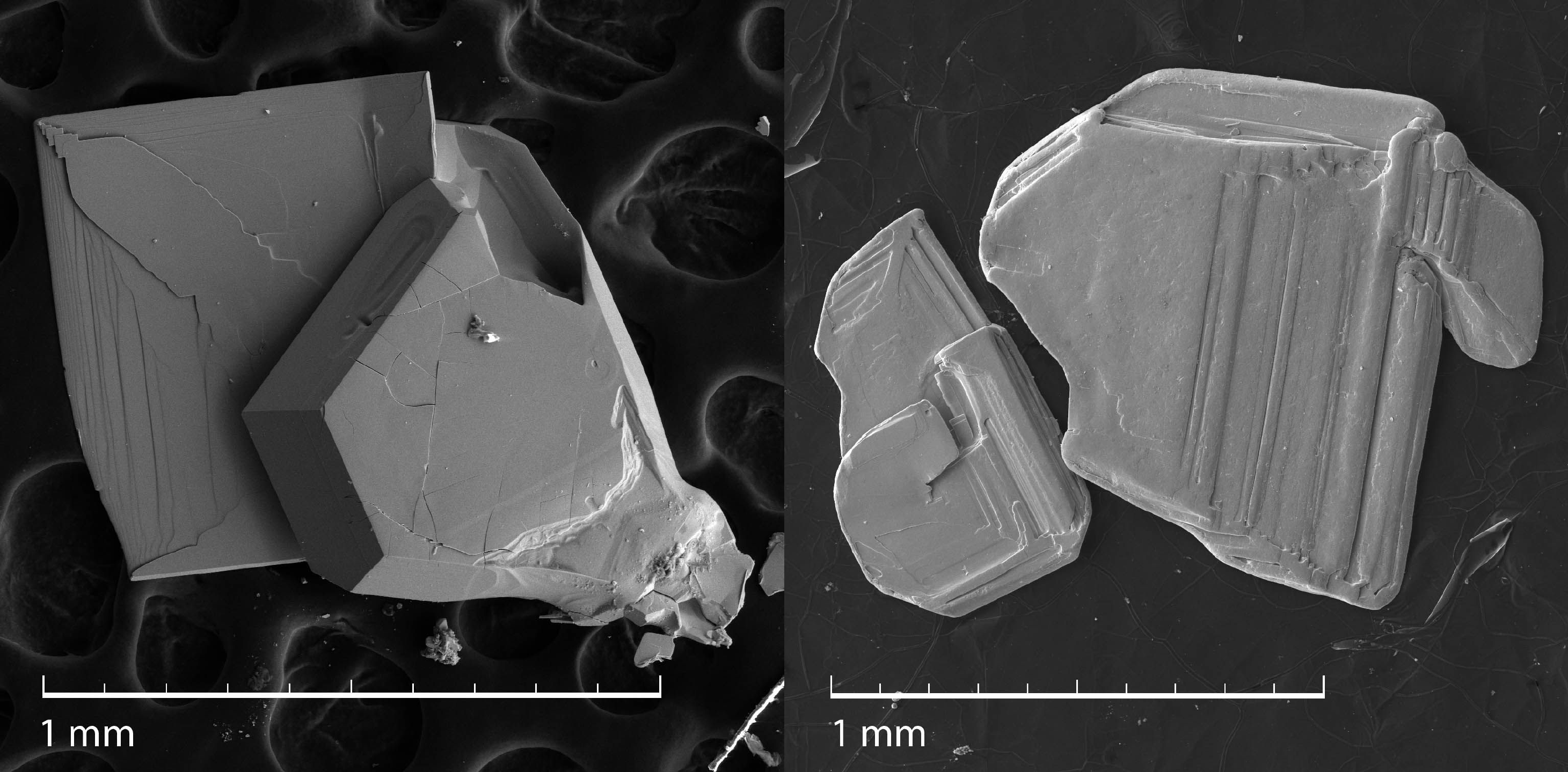}
  \caption{Scanning electron microscopy image of Fe(Se,Te) crystals with different morphology: hexagonal and tetragonal from x=0.21 batch (left), and tetragonal from x=0.17 batch (right). 
}
  \label{fgr:fig-SA1}
\end{figure}

In addition to the main batches of samples, second series of FeSe${}_{1-x}$Te${}_{x}$ crystals was also prepared at the same temperature conditions by a gas transport method using AlCl${}_{3}$ as a transport reagent\cite{bohmer2016variation}. Eventually, the pure tetragonal crystals, suitable for further investigation, were found only in three batches, prepared by the first method (referred to as the $x=0.17$, $x=0.21$,  and $x=0.25$ batches according to Te content in tetragonal crystals).  Crystals of hexagonal morphology (see Fig.\ref{fgr:fig0}, left panel) were also observed in all batches. Using an optical microscope, the thin tetragonal plate-like crystals up to 1 mm in size (see Fig.\ref{fgr:figSA1}, right panel) were selected and investigated by energy-dispersive spectroscopy and X-ray diffraction.
The initial chemical composition of batches, chemical composition of crystals obtained and their crystallographic characteristics along with the properties of pure FeSe are listed in Table~\ref{tbl:T1}.

At room temperature, X-ray unit cell determination of FeSe${}_{1-x}$Te${}_{x}$ samples with $x=0.17$,  $0.21$, and $0.25$ confirmed their belonging to the tetragonal modification. The values of lattice constants and volumes are close to those that were observed by Zhuang et al. for so-called phase $B$ in thin films of \fetese \cite{zhuang2014}. In that work, it was reported coexistence of two different tetragonal phases ($A$, $B$). Both phases were crystallized in the same tetragonal $P4/nmm$ space group and differed only by the unit cell volume. The phase with lower cell volume, denoted as $B$, shown to be stable up to $x=0.2$. The cell volumes of our samples FeSe${}_{0.83}$Te${}_{0.17}$ and FeSe${}_{0.79}$Te${}_{0.21}$ are 81.37 and 82.9~\r{A}$^{3}$, respectively, which almost coincides with the cell volume of  B phase in the thin films with the corresponding nominal compositions\cite{zhuang2014}. Our results show that both $c$ parameter and cell volume of FeSe${}_{1-x}$Te${}_{x}$ are increasing with the increase of tellurium content while the $a$ parameter remains almost constant in the studied range of compositions.

To prove the quality of the crystals, the study of magnetization was used. The crystals with hexagonal and tetragonal symmetries  are different by their magnetic properties. While the hexagonal phase is magnetic, \cite{tsuya1954, terzieff1978,prokes2015}, in the tetragonal crystals iron is nonmagnetic. The corresponding tetragonal phase is a Pauli paramagnet with a low and weakly dependent on temperature susceptibility. It helps to separate the crystals with the homogeneous tetragonal phase.


Magnetization was measured using a Quantum Design MPMS SQUID in a magnetic field of 10 Oe under
zero-field cooled conditions and 10 kOe for the $\chi (T)$ dependence in the temperature range 10-300~K.
Magnetic measurements were performed for the samples consisting of several (10-20) crystals for each of the batches. The crystals $ab$ planes were roughly (``by eye'') oriented  to be parallel to
the direction of magnetic field. The temperature dependence of magnetic susceptibility for crystals with tetragonal symmetry and the samples containing both crystals with tetragonal and hexagonal symmetry are presented in Fig.\ref{fgr:fig-2}. The susceptibility of the samples containing the crystals with hexagonal symmetry is divided by 30 to show all details in one graph. The $\chi (T)$ dependences for the samples with tetragonal only crystals are very similar to those of pure FeSe and Fe(SeS) compositions. They show a slow and almost linear increase over a wide temperature range. For these samples, the $\chi (T)$ slope changes the sign at temperatures below 25~K which indicates the presence of paramagnetic impurities. It is more pronounced for the $x=0.17$ sample where $\chi $ increases by approximately 1$\times$10$^{-6}$ emu/g$\cdot$Oe at 4 K. To provide the observed increase of $\chi (T)$, the concentration of the free Fe ions (the most probable source of the magnetic moment in the supposed paramagnetic impurities) in this sample was estimated to be only 5$\times$10$^{-4}$~mol$^{-1}$. This clearly indicates that the investigated crystals from $x=0.17$, $x=0.21$, and $x=0.25$  batches are composed almost entirely of the tetragonal phases of FeSe${}_{1-x}$Te${}_{x}$ and have a negligible amount of magnetic impurities.

\begin{figure}[h]
\centering
  \includegraphics[scale=0.5]{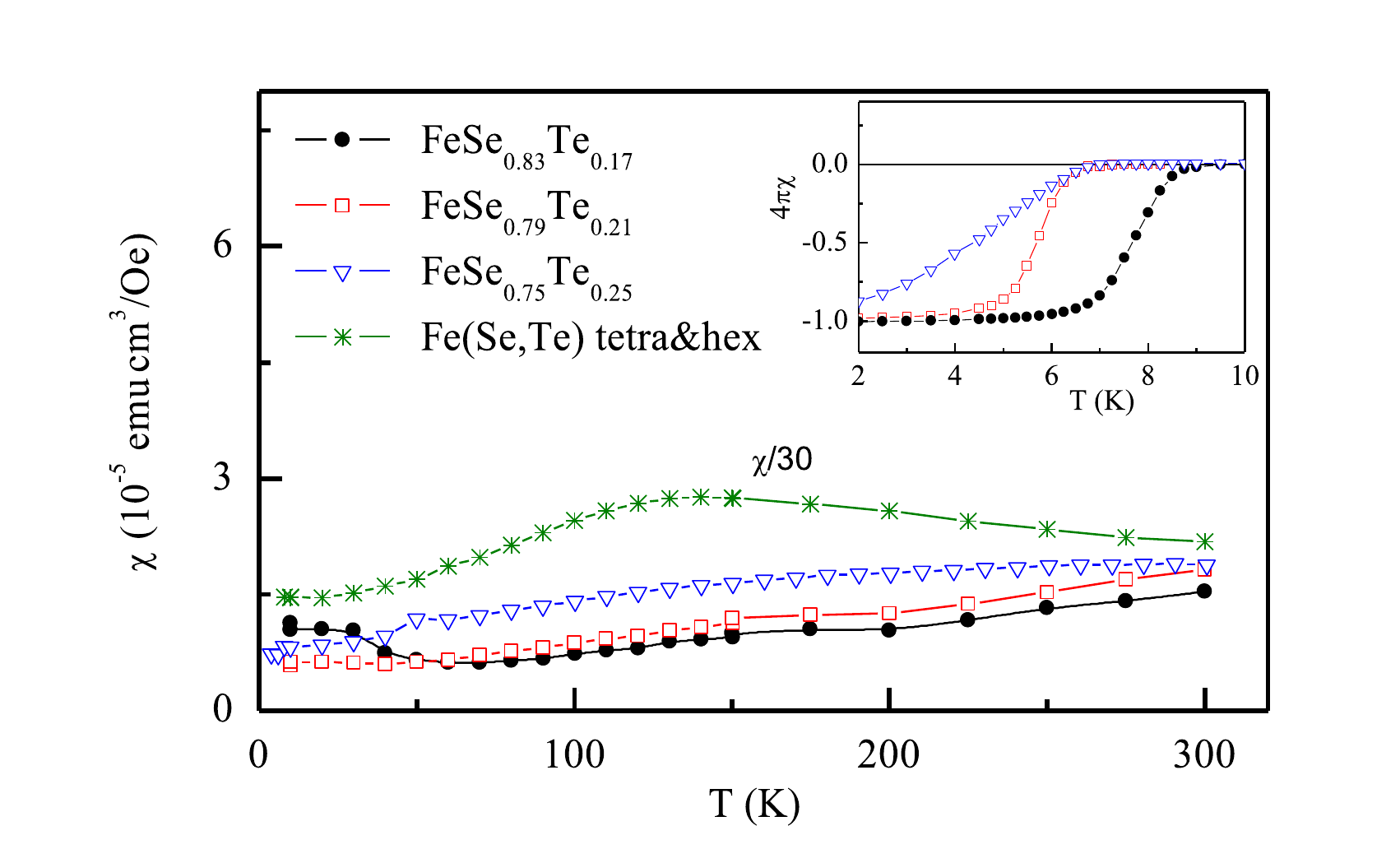}
  \caption{Temperature dependence of the susceptibility of tetragonal crystals from x=0.17, x=0.21, and x=0.25 batches along with the susceptibility of the sample with hexagonal crystals from x=0.21 batch. The susceptibility of the sample with hexagonal crystals is divided by 30. The inset shows the low-temperature ZFC susceptibility of the samples with tetragonal crystal structure.}
  \label{fgr:fig-SA2}
\end{figure}

The inset of Fig.\ref{fgr:fig-SA2} shows the zero field cooling (ZFC) $\chi (T)$ measured at 10 Oe. For the samples $x=0.17$ and $x=0.21$ the curves demonstrate a sharp superconducting transition at approximately 6 and 8 K with full Meissner shielding.

\begin{figure}[ht]
\includegraphics[scale=0.5,angle=0]{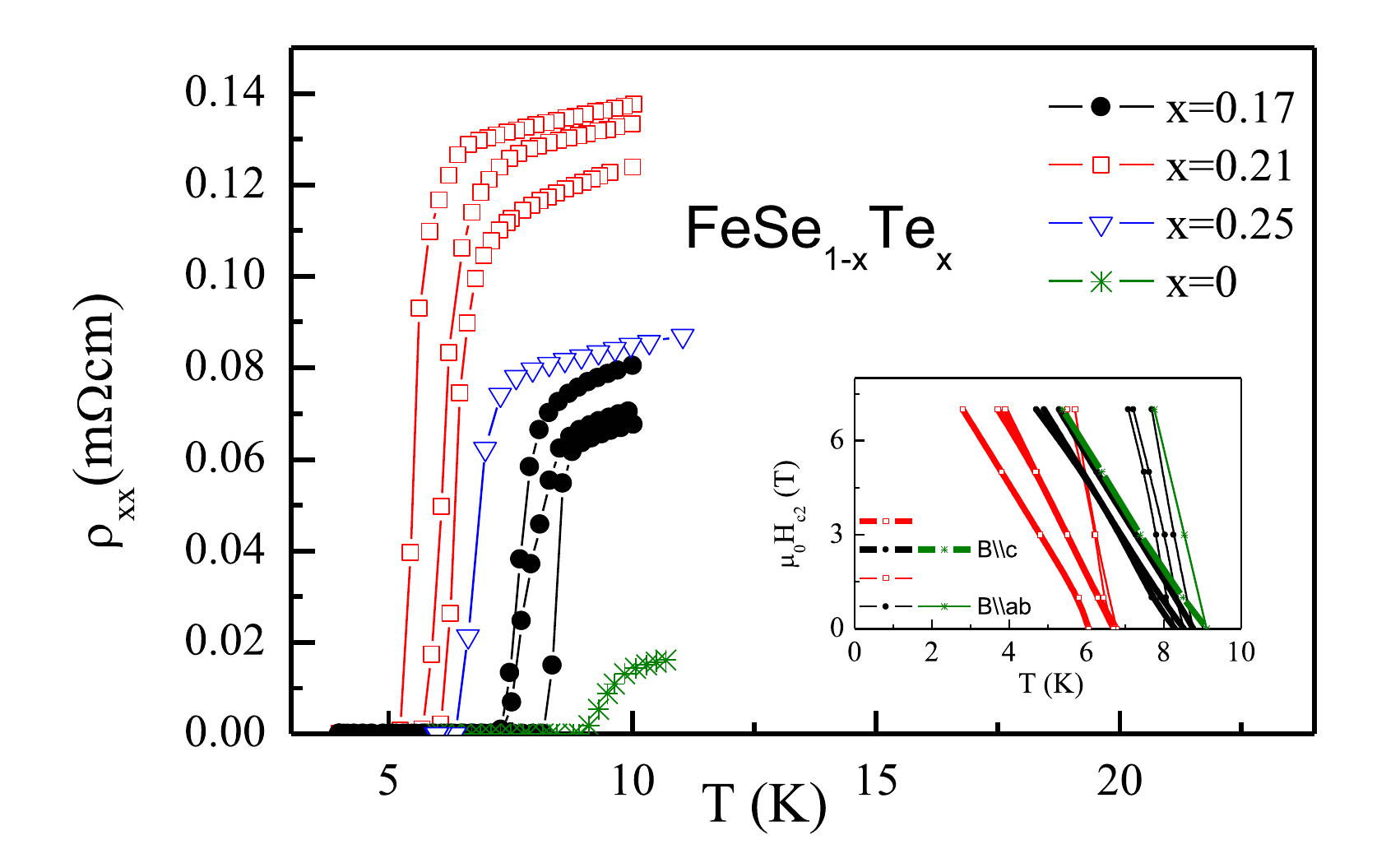}
\caption{ Temperature dependence of the resistivity $\rho_{ xx}$ at low temperatures for some crystals from each of x=0.17, x=0.21 and x=0.25 batches and for the reference FeSe crystal. Inset: Temperature dependence of$H_{C2}$ critical field for B//c (thick lines) and B//ab (thin line). }
\label{fgr:fig_SA3}
\end{figure}

\begin{figure}[t]
\centering
  \includegraphics[scale=0.5]{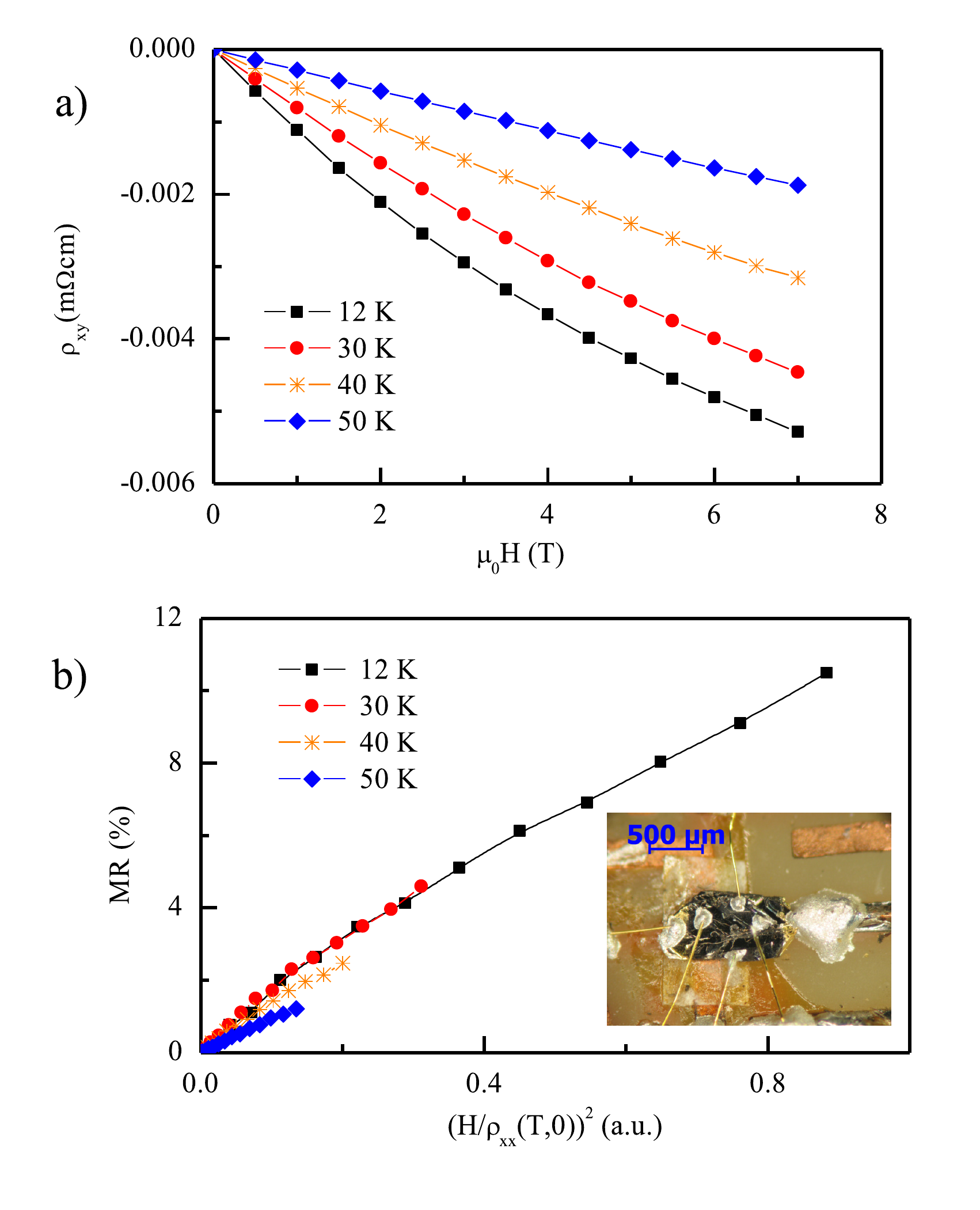}
  \caption{ Magnetic field dependence of $\rho_{xy}$ (a) and MR versus $(B/\rho_{xx}(0))^{2}$ (Kohler plot) (b) at temperatures between 12 and 50 K and magnetic field up to 7 T for FeSe${}_{0.83}$Te${}_{0.17}$. Inset: The optic image of the studied FeSe${}_{0.83}$Te${}_{0.17}$ crystal.}
  \label{fgr:fig-SA4}
\end{figure}

 We studied the transport properties only for those crystals whose quality was confirmed by magnetic measurements. Besides, for each of $x=0.17$ and $x=0.21$ batches three different crystals were measured to verify the homogeneity of the batches. Electrical measurements were done on cleaved samples with contacts made by sputtering of Au/Ti layers through a precisely machined mechanical mask which minimizes relative errors of $xx$ and $xy$ components of the measured resistivity tensor. The crystal sizes were carefully measured by an optical microscope to accurately determine the resistivity values.

The Fig.\ref{fgr:fig_SA2} shows $\rho_{xx}(T)$ dependencies around the superconducting transition temperature. The transitions are narrow and the difference in the values of $T_{c}$ for one batch is substantially less than the difference between batches. This again shows a good homogeneity of properties of crystals in batches.

To further study the quality of crystals, we measured the field dependence of $T_{c}$ for a field oriented perpendicular and parallel to the surface of the plate-like sample but perpendicular to a current direction in both cases. For a polycrystalline sample, no significant difference is expected. For the samples studied, $H{}_{C2}(T)$ for $H//c$ and $H//ab$ differ significantly  (see the inset in Fig.\ref{fgr:fig_SA3}) and the ratio $(d\mu_{0}H_{c2}/dT)^{H//ab}/(d\mu_{0}H_{c2}/dT)^{H//c}$ is in the range 3-3.5.
The observed anisotropies ensure that the crystals are well oriented, at least along $c$-axis. The X-ray diffractometry analysis reveals a mosaic structure in the $ab$ plane for the most of the crystals studied. Nevertheless, the tetragonal crystals, well-oriented along the $c$ axis, allow to measure correctly the resistivity tensor components in the $ab$ plane and to extract the parameters of carriers using a multicarrier analysis.

In the Fig.~\ref{fgr:fig-SA4} we present the magnetic field dependence of the resistivity tensor components  for $x=0.17$. The remarkable feature of the low-temperature $\rho_{xy}(B)$, shown in Fig.\ref{fgr:fig-5}a, is a concave upward shape. Similar concave upward $\rho_{xy}(B)$ dependencies is well known for low-temperature orthorombic phases of pure FeSe and BaFe$_{2}$As$_{2}$ compounds and can be explained by the presence of the highly mobile electron component.

The Fig.\ref{fgr:fig-SA4}b shows the MR=($\rho_{xx}(B)$-$\rho_{xx}(0)$)/$\rho_{xx}(0)$ plotted versus $(B/\rho_{xx}(0))^{2}$ (Kohler plot) for $x=0.17$. This plot shows that the Kohler's rule is violated, which in particular may be explained by a Lifshitz transition accompanying a supposed  nematic transition in these compounds.

\subsection{The carriers parameters extraction}

 For tetragonal crystals:
\begin{eqnarray}
\sigma_{xx}=\sigma_{yy}=\frac{\rho_{xx}}{(\rho_{xx}^{2}+\rho_{xy}^{2})} \nonumber\\
-\sigma_{xy}=\sigma_{yx}=\frac{\rho_{xy}}{(\rho_{xx}^{2}+\rho_{xy}^{2})}
\end{eqnarray}

where $\sigma_{ij}$ are the conductivity tensor components and $\rho_{ij}$ are the resistivity tensor components

The conductivities of bands are additive, and within the relaxation-time approximation for an arbitrary number of bands, we can write:
\begin{eqnarray}
\sigma_{xx}=F_{R}(B)\equiv\sum_{i=1}^{l}\frac{|\sigma_{i}|}{(1+\mu_{i}^{2}B^2{})}\nonumber\\
\sigma_{xy}=F_{H}(B)\equiv\sum_{i=1}^{l}\frac{\sigma_{i}\mu_{i}B}{(1+\mu_{i}^{2}B^2{})}\nonumber\\
\sigma_{i}=en_{i}\mu_{i}
\end{eqnarray}

where $i$ is the band index, $e$ is the electron charge, $\sigma_{i}$ is the conductivity at $B=0$, $\mu_{i}$ is the mobility, $n_{i}$ is the carrier concentration, and $l$ is the number of bands.

 To determine $\mu_{i}$ and $n_{i}$ we minimize the residual $\phi$:
\begin{equation}
\phi=\frac{1}{N}\sum_{k=1}^{N}{\bigg[\bigg({\frac{\sigma_{xx}[k]-  F_{R}(B[k])}{\sigma_{xx}[k]}\bigg)}^{2}+\bigg({\frac{\sigma_{xy}[k]- F_{H}(B[k])}{\sigma_{xy}[k]}\bigg)}^{2} \bigg]} \nonumber
\end{equation}
where $\sigma_{xx}[k]$, $\sigma_{xy}[k]$, and $B[k]$ are the values of $\sigma_{xx}$, $\sigma_{xy}$, and $B$ at experimental point $k$, and $N$ is the number of the measured points.

\begin{figure}[ht]
\includegraphics[scale=0.5,angle=0]{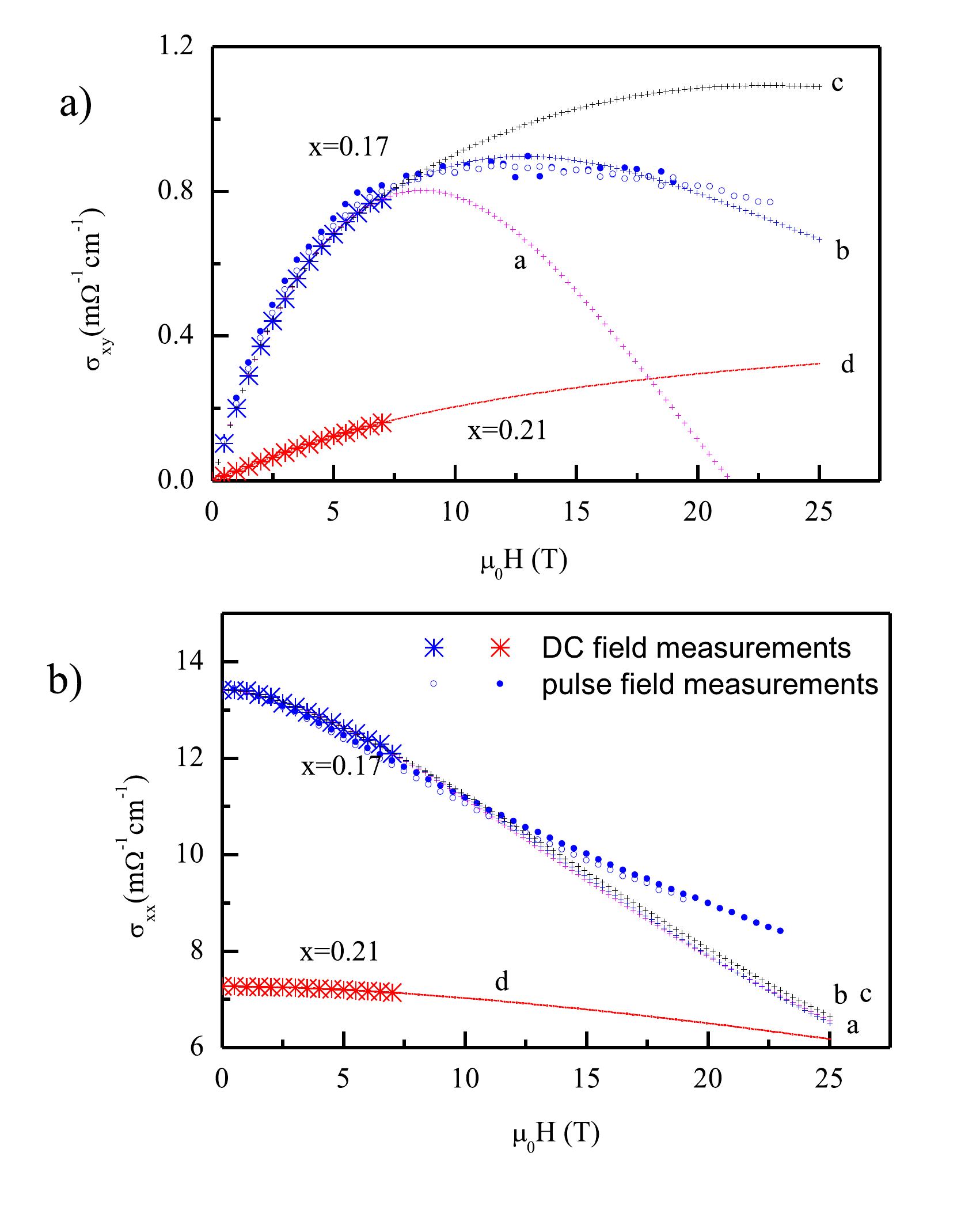}
\caption{  Experimental data at 12 K and the simulation of the fitting results for $\sigma_{xy}$ (a) and $\sigma_{xx}$ (b). For DC data in field up to 7 T, the curve "a" is the best fit for FeSe${}_{0.83}$Te${}_{0.17}$ sample , the curve "b" is the fit with an almost equal electron and hole concentrations in main bands, and the curve "c" is the fit with an almost equal electron and hole mobilities in main bands. The curve "d" is the best fit for FeSe${}_{0.79}$Te${}_{0.21}$ sample,. }
\label{fgr:fig_SA5}
\end{figure}

Results of the  multicarrier analysis are listed in Table \ref{tbl:TS1}. The field range up to 7 T does not allow the accurate determination of the parameters of the main bands because of the carriers mobility in \fetese is low. The curves ``a'', ``b'' and ``c'' in Fig.\ref{fgr:fig_SA5} corresponding to parameters listed in Table \ref{tbl:TS1} almost perfectly reproduce the DC experimental data in 7 T range. The pulse field data improve the accuracy of determining the parameters, but strictly speaking, this data also not allow us to confidently assert that the concentration of electrons has exceeded the concentration of holes. On the other hand our magnetotransport data well agree with this supposition. 

\begin{table*}[ht]
\small
  \caption{\ The results of the simultaneous fitting of $\sigma_{xy}(B)$ and $\sigma_{xx}(B)$ using the three-band model in field range up to 7 T  for FeSe${}_{0.83}$Te${}_{0.17}$ sample (``$a$'', ``$b$'', and ``$c$'' ), for the FeSe${}_{0.79}$Te${}_{0.21}$ sample (``$d$'') and in field range up to 25 T for FeSe${}_{0.83}$Te${}_{0.17}$ sample (HF) }
  \label{tbl:TS1}
  \begin{tabular*}{\textwidth}{@{\extracolsep{\fill}}llllllll}
    \hline
    Fit &  $n_{h}$ & $\mu_{h}$  &  $n_{e1}$ & $\mu_{e1}$ &  $n_{e2}$ & $\mu_{e2}$ & $\phi$ \\
			&(10$^{19}$ cm$^{-3}$) & (cm$^{2}$/Vs) & (10$^{19}$ cm$^{-3}$) & (cm$^{2}$/Vs)&(10$^{19}$ cm$^{-3}$) & (cm$^{2}$/Vs) &  \\
    \hline
    a& 14.8 &305 &7.2 & 539 & 0.06 & 2809&    4.9 10$^{-6}$\\
    b& 10.2 &360 &10.2 & 431 & 0.15 & 2207&    1.5 10$^{-5}$\\
    c& 8.8 &385 &11.9 & 385 & 0.19 & 2056&    2.1 10$^{-5}$\\
   d & 13.3 &152 & 14.7 & 163 & 0.11 & 890 & \\
    HF& 11.5 &303 &11.9 & 343 & 0.32 & 1830&    2.0 10$^{-4}$\\

    \hline
  \end{tabular*}
\end{table*}

\begin{figure}[ht]
\includegraphics[scale=0.5,angle=0]{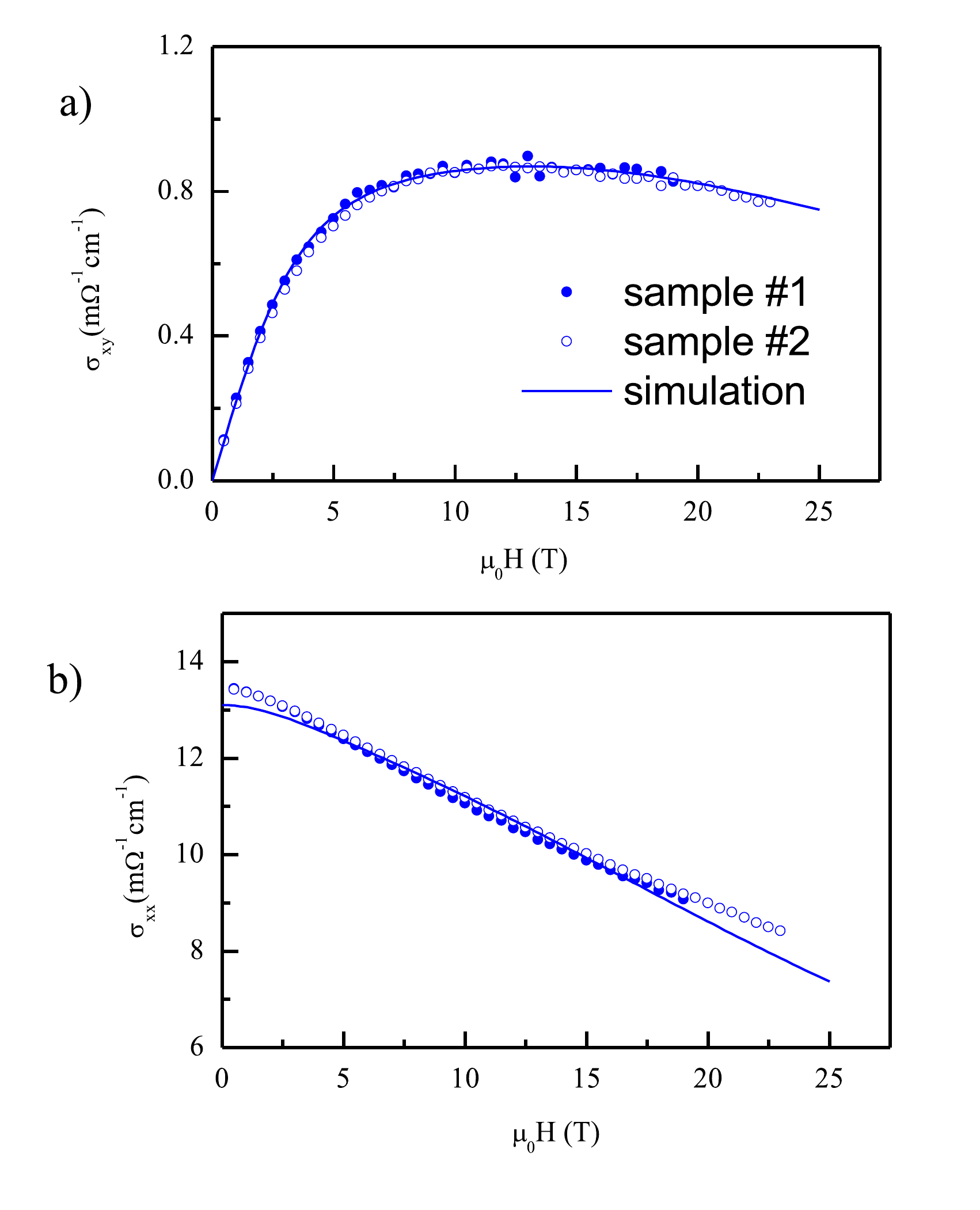}
\caption{ Experimental data in pulse field and the best three-band  fit for FeSe${}_{0.83}$Te${}_{0.17}$. }
\label{fgr:fig_SA6}
\end{figure}

The Fig.\ref{fgr:fig_SA6} shows the experimental data in pulse field together with the three band fitting result that is marked in Table \ref{tbl:TS1} as ``HF''. The sample \#1 is the same one for which we give the data in DC 7 T field range. It is worth to note that $\rho(B)$ curves in DC and pulse field are always slightly differ for the same samples, probably due to the behavior of nematic domains.

\end{document}